\documentclass[pra,twocolumn,showpacs,superscriptaddress]{revtex4-1}

\usepackage{ulem}
\usepackage{graphicx}
\usepackage{amsmath}
\usepackage{amssymb}
\usepackage{latexsym}
\usepackage{fancyhdr}
\usepackage{array}
\usepackage{amsfonts}
\usepackage{mathrsfs}
\usepackage{mathtools}
\usepackage{color}
\usepackage{verbatim}
\usepackage{physics}
\usepackage[caption=false]{subfig}
\usepackage{natbib}
\usepackage{enumitem}
\usepackage{bbold}
\usepackage[colorlinks=true,urlcolor=blue,citecolor=blue,linkcolor=blue]{hyperref}

\usepackage{float}

\newcommand{\marcin}[1]{{\color{black} #1}}
\newcommand{\marcinN}[1]{{\color{black} #1}}
\newcommand{\marek}[1]{{\color{black} #1}}
\newcommand{\marcinA}[1]{{\color{black} #1}}

\begin{document}
\title{Physics and Metaphysics  of Wigner's Friends: Even performed pre-measurements have no results}

\author{Marek \.Zukowski and Marcin Markiewicz}
\affiliation{International Centre for Theory of Quantum Technologies (ICTQT),
University of Gdansk, 80-308 Gdansk, Poland}

\date{\today}

\begin{abstract}
``The unambiguous account of proper quantum phenomena must, in principle, include a description of all relevant features of experimental arrangement" (Bohr). The measurement process is  composed of pre-measurement (quantum correlation of the system with the pointer variable), and an  irreversible decoherence  via interaction with an environment. The system ends up in a  probabilistic mixture of the eigenstates of the measured observable.  For the pre-measurement stage,  any attempt to introduce an `outcome’  leads, as we show, to a logical contradiction, $1=i$.  This nullifies claims that a modified concept of Wigner’s Friend, who just pre-measures, can lead to valid results concerning quantum theory. 
\end{abstract}

\maketitle

Wigner's Friend is back in the discussion about the foundations of quantum mechanics.
It is used in the Frauchiger-Renner gedankenexperiment \cite{FR-RN} aimed at showing that ``quantum theory cannot consistently describe the use of itself". The crucial issue of the gedankenexperiment is the undefined status of the measurements  by the Friends, which can be interpreted either, as  proper irreversible measurements or, as  (in principle) reversible \textit{pre-measurements}  with no (irreversible) decoherence.

Many authors  discuss consequences of the  paradox of \cite{FR-RN} for various interpretations of quantum mechanics. Such a discussion is in \cite{FR-RN}, especially in the initial version. Some  indicate that the  paradox is due to  referring to incompatible experimental contexts \cite{DREZET18, FORTIN19, LOSADA19}, others state that decoherence must be involved in the discussion of a measurement process \cite{RELANO1, RELANO2}.
{\color{black} Additional hidden assumptions in the  paradox of Ref. \cite{FR-RN} have also been suggested \cite{HEALEY18, SUDBERY19}.}
There are comments suggesting that the reason for the paradox is the indefiniteness of the \textit{Heisenberg cut} \cite{LALOE18, TAUSK}, and insufficient objectivity of outcomes of a measurement \cite{SUAREZ, WAAIJER}. A discussion concerning "stable vs. relative" facts can be found in \cite {ROVELLI}. {The paradox has also been discussed in the context of QBist \cite{CAVALCANTI20, DEBROTA20} and Everett's interpretations \cite{BUB19}}. We agree with those authors who underline complementarity and need to take into account decoherence, as an indispensable part of the measurement process.

We show that the paradox can be   nullified in  a three-step  argumentation.

{\it Step one}. We show, in concurrence with \cite{RELANO1, RELANO2}, that if Friends perform proper measurements, and universal validity of the quantum description is taken into account, the gedankenexperiment presented in \cite{FR-RN}   never leads to  those correlations of results that are the basis of the  contradiction presented in \cite{FR-RN}.  Universal validity of  quantum description must take into account measurement devices and the measurement process. Usually one skips this, but it is essential, at least in a sketchy way, in the case of {\it sequential} measurements.

The basis of \marcin{the step} is the quantum theory of measurement. It shows how pre-measurement and decoherence lead, via a unitary dynamics, to a density matrix for the system that is a classical probabilistic mixture of eigenstates of the measured observable. 
 See e.g. \cite{BOHM, BUSCH}, or the introduction in \cite{HAAKE}. The most  definitive  exposure of it is in the trailblazing works of Zurek \cite{ZUREK1, ZUREK2}.
The crucial stages \marcin{of the measurement process} are as follows.

{\it Pre-measurement}: the measured system, $s$, say in a state $|\Psi\rangle_s=\sum_i\alpha_i|\Psi_i\rangle_s $, where $|\Psi_i\rangle$ are eigenstates of the measured observable, gets entangled (correlated) with the pointer observable of the measuring device (states of the preferred pointer  basis of it are denoted here as $\ket{P_i}_p$).  A unitary evolution leads from
$\sum_i\alpha_i|\Psi_i\rangle_s \ket{P_{initial}}_p$ to $\sum_i\alpha_i|\Psi_i\rangle_s \ket{P_i}_p$. Pre-measurements are in principle reversible (for  illustrations of this see  \cite{FEYNMAN}, where a  different wording is used).

{\it Decoherence}: the  measuring device is a macroscopic object, thus it has its own internal and external {\it uncontrollable} environments, which can be thought of  as  zillions of microscopic degrees of freedom which we must ignore in the description. This is because: (i) we are interested only in the position of the pointer, (ii) like in thermodynamics it is impossible to describe the evolution of the micro-states of these degrees of freedom. 
Unitary interaction of the pointer variable  with these environments leads to irreversible decoherence,  \cite{ZUREK1, ZUREK2} (because we cannot control all these degrees of freedom of the environment), and to a broadcast of the results, see e.g. \cite{OLLIVIER, KORBICZ}. After the decoherence the effective system-pointer state 
is a classical   mixture of states $|\Psi_i\rangle_s  \ket{P_i}_p$ with respective probabilities given by  $|\alpha_i|^2$.

The contradiction shown in \cite{FR-RN} is a consequence of equating, in the case of Friends, of  pre-measurement with measurement. $\square$

{\it Step two}. We give the paradox a second chance and  equate  pre-measurement with measurement.  Friend just makes pre-measurements (we call such Friend `virtual'), and `super Wigner' is able to make measurements in a basis in which the Friend (and her lab) are in a superposition of her seeing one or the other of mutually exclusive results (this is the cornerstone of the gendanken experiment of \cite{FR-RN}). We show that the action of Wigner precludes any operational meaning of  results of virtual Friends. The `outcomes' of Friends are thus counterfactuals. {Our argumentation is closely related to the quantum-marker quantum eraser experiments, see \cite{KWIAT1, KWIAT2, KWIAT3} {and in Appendix \ref{ApD}}, and is concurrent with those papers which pointed that complementarity nullifies the paradox, e.g. \cite{DREZET18, FORTIN19, LOSADA19}} $\square$

{\it Step three}. We allow  for theoretical description that includes such  counterfactual results by virtual Friends (as it is  in \cite{FR-RN}), and assume that their mutual algebraic relations must be concurrent with quantum mechanical predictions.
We present a {\it reductio ad absurdum} reasoning based on Greenberger-Horne-Zeilinger-type (GHZ) correlations \cite{GHZ}, showing that a (virtual) Friend who performs only a pre-measurement cannot ascribe to this process any counterfactual  notion of an 'outcome', as this  leads to a  contradiction. This finally nullifies the paradox.
Step three is our main result.$\square$

This also implies  that the related gedankenexperiment of Brukner: "A no-go theorem for observer-independent facts" \cite{BRUKNER, EXP1, EXP2}  does not imply that measurement outcomes can be observer-dependent. Instead, it is a form of refutation of counterfactual pre-measurement outcomes, which involves additional assumptions of Bell-type locality, and freedom of choice of measurement settings by observers. Our reasoning does not rest on these. Outcomes of pre-measurements are not merely counterfactual, they are an internally inconsistent notion (and therefore are not `facts').  As such they are useless as  a part of a  premise of a logical implication aimed at deriving true statements. 
 
Before we proceed, we must point to the following.
In the gedankenexperiments of both papers \cite{FR-RN} and \cite{BRUKNER} it is assumed that  Friends and Wigners are in a different relation than the one assumed by {E.P. Wigner} (we add initials  `E.P.' whenever referring to the Author of \cite{WIGNER}).
In three of the situations in  Ref.  \cite{BRUKNER}, and in the situation studied by \cite{FR-RN}, the authors confront the `results' of fixed-setting (pre)-measurements done by (virtual) Friends, that are followed by final measurements  by (super) Wigners, on the compound system of (virtual) Friend, his measuring device and the system $s$ in question, which are effectively complementary (or partially complementary) to the pre-measurements by Friends. We call a measurement by (super) Wigner `effectively' complementary, if after it there is no way to establish what could have been the initial state of the system, had it been  an unknown  eigenstate  of Friends (pre-)measurement, i.e. whenever $\ket{\Psi}_s=\ket{\Psi_i}_s$.  

This is absolutely not in concurrence with E.P. Wigner's ideas, who in
 Ref. \cite{WIGNER},  assumed that whenever Wigner `measures', then it is in a basis which is in concurrence with the measurement of Friend, and he just confirms Friend's result. 
We quote: {\it "my friend" has the same types of impressions and sensations as I - in particular he is not in that state of suspended animation which corresponds to the wave function $\alpha (\Psi_1 \times X_1)+ \beta(\Psi_2 \times X_2),$ } \cite{WIGNER}. In E.P. Wigner's notation $\Psi_i$ are mutually orthogonal  wavefunctions of the measured system, and $X_i$ are mutually orthogonal states of the Friend or her consciousness.
This is why we call
the `agents' used in both Refs. \cite{FR-RN} and \cite{BRUKNER, EXP1, EXP2}, see also ref. \cite{DEUTSCH}, {\it super-Wigners and virtual-Friends}.


{\bf Step one.} We  present the gist of our argumentation: if Friends perform full measurements the paradox of Ref. \cite{FR-RN} is void (detailed analysis which uses the peculiarities of the configuration of \cite{FR-RN} is in  Appendix, Section \ref{ApA}). The same holds for the (gedanken) experiments of \cite{BRUKNER, EXP1, EXP2} (no violation of the Bell-like inequalities).

We consider a  Friend $F_g$ (non-virtual!), who performs a {\it measurement} on a system $g$ in her isolated lab $L_g$ using a device
$D_g$. Pure states of the subsystems will have an appropriate subscript. Assume that the initial state of the system is  ${\alpha}\ket{H}_g+\beta\ket{T}_g$, with $\alpha\beta\neq 0$, and $\ket{T}$ and $\ket{H}$ forming a basis. A pre-measurement by Friend in this  basis leads to a correlated state  ${\alpha}\ket{H}_g\ket{H}_{D_g}\ket{H}_{F_g}+\beta\ket{T}_g\ket{T}_{D_g}\ket{T}_{F_g}$. States `$H$' and `$T$' for other subsystems are also assumed to be normalized and orthogonal.

{\it Super-Wigner} $W_g$: an agent who is capable to perform measurement in bases containing 
\begin{equation} \label{BASIS}
\ket{\pm; \phi}_{L_g} =\frac{1}{\sqrt{2}}\big(\ket{H}_{L_g}
\pm e^{i\phi}\ket{T}_{L_g}\big),
\end{equation}
or more general superpositions of $\ket{X}_{L_g}=\ket{X}_g\ket{X}_{D_g}\ket{X}_{F_g}$, where $X=H,T$.
Super-Wigners perform full (irreversible) measurements, which involve decoherence.  

This is daunting task, as especially the device, $D_g$, and the Friend $F_g$ may be complex objects of many degrees of freedom. Still, as long as all these degrees of freedom are controllable, a Super-Wigner might be able to perform such an experiment. 

We now shall consider the consequences of using the quantum measurement theory, which involves decoherence and irreversibility of recorded results, for  Friends and Wigners performing full proper measurements.

Let us consider Friend's lab $L_g$.
 The environment in the lab, $E_g$, causing the decoherence, might be in a mixed initial state, but for simplicity of presentation we put it  as a pure one.
The two (distinguishable, according to the measurement theory) states of the environment related with distinct states of the pointer-device-Friend \marcin{system} $X=H, T$
will be denoted as $\ket{X}_{E_g}$.
 The state of \marcin{the entire lab consisting of} $g\otimes D_g\otimes F_g\otimes E_g$  after the full measurement process, is given by:
\begin{equation} 
\alpha\ket{H}_{L_g}\ket{H}_{E_g}
+\beta\ket{T}_{L_g}\ket{T}_{E_g}.
\end{equation}
$W_g$ in his following measurement, despite being a super-Wigner, cannot undo the irreversible interaction with the environment (and thus cannot undo the records). 
 Because of the entanglement with environment $E_g$, he effectively deals with a reduced density matrix of $g\otimes D_g\otimes F_g$. 
Therefore measurements by a super-Wigner in bases containing $\ket{\pm, \phi}$ give just random results.

Thus, \marcin{if one interprets the actions by Friends as (irreversible) proper measurements}, a consistent use of the measurement theory nullifies the paradox.  $F_g$ records either $H$ or $T$. If it is $T$, there is an isomorphism with situation 2, discussed by us in Appendix (Section \ref{Apa1}), and there is no way to formulate the paradox. Also in the case of the Bell-like inequality of Brukner \cite{BRUKNER, EXP1, EXP2}, full measurements by Friends preclude its violation (see Appendix, Section \ref{ApC}). 

{\bf Step two.} 
Consider Friends who make only pre-measurements, no uncontrollable environment in their labs.

{\it Virtual-Friend} $F_g$: an agent who is capable to perform a reversible pre-measurement (virtual-measurement). She does  not interact with an uncontrollable environment, and is able to steer herself and the lab into a superposition $\alpha \ket{H}_g\ket{H}_{D_g}\ket{H}_{F_g} +\beta\ket{T}_g\ket{T}_{D_g}\ket{T}_{F_g}$, in response to the  system $g$ being initially in superposition $\alpha \ket{H}_g+\beta\ket{T}_g$.  Virtual Friends face no decoherence. However, somehow they associate with their pre-measurement a concrete result, $H$ or $T$.

\marek{Virtual Friends' actions can be undone, e.g. by a suitable measurement by a Super-Wigner, which erases them, or by an inverse unitary process which would uncouple the system $g$ from the rest of the lab and restore its state (both possible if there are no uncontrollable couplings with environment $E_g$).}


\marcin{We deal exactly with such  virtual Friends and Super-Wigners in \cite{FR-RN} and   \cite{BRUKNER, EXP1, EXP2}.} 
 Super-Wigners erase the `results' by their Friends, if their measurement bases are like in \cite{FR-RN}, or more generally not concurrent with the pre-measurement basis  of their virtual Friend. Say, $W_g$ performs a measurement which contains basis states \marcin{$\ket{\pm,\phi=0}_{L_g}$},  a result which is concurrent with $\ket{-,0}_{L_g}$. 
Then, 
all that he knows is that the `result' of $F_g$ is either $H$ or $T$ with probability $1/2$, i.e., nothing. This is an instance of an  effectively complementary (unbiased) measurement with respect to the pre-measurement by the virtual-Friend.

Further, there is no way to recover the hypothetical  `result' of $F_g$ as the measurement of $W_g$ involves irreversible decoherence. In Ref. \cite{FR-RN} the results of Super-Wigners  halt the `experimental procedure', or restart it, thus they are accessible to a macroscopic  experimenter or an  automaton controlling the experiment. The `results' by virtual-Friends are inaccessible,  erased after an effectively complementary measurement by super-Wigner.  The pre-measurement results are thus {\it counterfactual}, because they could have been read out, had Wigner measured in a basis concurrent with the pre-measurement basis of the Friend, namely the one which contains states $\ket{X}_g\ket{X}_{D_g}\ket{X}_{F_g}$ . But he does not in \cite{FR-RN}, neither  does at least one of the Wigners in three out of four situations covered by the Bell-like scenario of \cite{BRUKNER}. 


Thus seemingly the argumentation  in both papers holds only for virtual Friends, who get counterfactual `outcomes'. But, even this is impossible. In both papers it is assumed that virtual Friends somehow get (can be aware of, or whatever) `outcomes' of their pre-measurements.  We show below a logical inconsistency of this assumption. This nullifies the paradoxes of \cite{FR-RN} and \cite{BRUKNER, EXP1, EXP2} in the case we interpret Friend's actions as reversible pre-measurements, not involving decoherence, but the nullification is at a deeper level. 

{\bf Step three.}
{\it Counterfactual outcomes of  pre-measurements, reductio ad absurdum---} 
Consider a \marek{three-stations GHZ-like experimental arrangement. In each station there is a qubit, say, encoded in states of an atom in a trap. The qubits are  in a GHZ state.  Such a state could be prepared an event ready way by an entanglement swapping experiment \cite{ENT-SWAP} in the version  of \cite{HARALD}, generalized to three stations, say with an adaptation of  the star configuration of \cite{ZZW}. }
 
    In each measurement station, numbered by $m=1,2$ and $3$, we have  a virtual Friend, $F_m$, and a super-Wigner, $W_m$ who operate in sequence. Assume that virtual Friends have a device which allows them to perform (only) pre-measurements. 
    Each $F_m$ makes this in  a fixed pre-measurement basis. Next   $W_m$,  measures  the joint system consisting of the  Friend and the system, $s_m$, in a basis which is effectively complementary to the one  used by  $F_m$.
    
    Virtual Friends somehow associate `outcomes' with their pre-measurements. These are the counterfactuals, defined in {\it step 2}. Note, that the `outcomes' are counterfactual, while pre-measurements are {\it real} actions, described by unitary transformations possible thanks to their devices.
    
    The initial state of the qubits  $\ket{GHZ}_{s_1s_2s_3}$ is given by
    \begin{eqnarray}\label{GHZ}
    \frac{1}{\sqrt{2}}\left(\otimes_{m=1}^3  \ket{0^{(1)}}_{s_m}
    +\otimes_{m=1}^3  \ket{1^{(1)}}_{s_m} \right),
    \end{eqnarray}
    where $\ket{l^{(n)}}_{s_m}$ is the $l$-th state of an orthonormal basis $n$ for  $s_m$. 
    
    In the following,  bases $n=1,2,3$ are  
    mutually unbiased (complementary). The elements of bases $n=2,3$ in terms of kets of basis $n=1$
    are given by $\ket{l^{(2)}}_{s_m}=\frac{1}{\sqrt{2}}\left(\ket{0^{(1)}}_{s_m}+e^{il\pi}\ket{1^{(1)}}_{s_m}\right)$ and  $\ket{l^{(3)}}_{s_m}=\frac{1}{\sqrt{2}}\left(\ket{0^{(1)}}_{s_m}+e^{i(\pi/2+l\pi)}\ket{1^{(1)}}_{s_m}\right)$. We define the eigenvalues of the related observables as $(-1)^{(l)}$. 
    
    In all stations local virtual-Friends pre-measure the qubit in   basis $\ket{l^{(3)}}_{s_m}$. 
    That is, the local Friend gets correlated with the local system due to the following  unitary process:
    $\ket{l^{(3)}}_{s_m}\ket{initial}_{F_m} \rightarrow \ket{l^{(3)}}_{s_m}\ket{l^{(3)}}_{F_m}\equiv \ket{l^{(3)}}_{s_mF_m}, $ where $\ket{l^{(n)}}_{F_m}$ are orthonormal. 
    Next  super-Wigner measures joint system $s_m\otimes F_m$. This is  done in  
    the effectively complementary basis $ \ket{l^{(2)}}_{s_mF_m}$.   
   
    {\color{black} Effectively complementary measurements by Wigners, have the following features.
  In the usual case in which we do not introduce Friends, the states of a system $\ket{l^{(1)}}$  are expressed in the following way  in terms of  the measurement basis states (for  setting $n=2$):
    \begin{eqnarray} \label{PSI}
   & \ket{l^{(1)}}= \sum_{j=1,2}\braket{j^{(2)}}{l^{(1)}}\ket{j^{(2)}}.&
    \end{eqnarray}
    In this paragraph we drop indices $m$, and $s_m$, however for system-Friend states retain a reduced subscript $sF$.
    The sequential action of a virtual-Friend, pre-measuring   in basis $n=3$, and next her super-Wigner measuring in basis $n=2$, which we consider here, can be put as follows
    \begin{eqnarray} \label{PSI2}
   & \ket{l^{(1)}}\ket{initial}_{F} \rightarrow\sum_{j=1,2}\braket{j^{(3)}}{l^{(1)}}\ket{j^{(3)}}\ket{j^{(3)}}_{F}& \nonumber \\
   & = \sum_{j=1,2}\braket{j^{(2)}}{l^{(1)}}\ket{j^{(2)}; W_1}_{sF},
    \end{eqnarray}
where $\rightarrow$ stands for a unitary evolution, $\ket{initial}_{F}$ is the initial state of the  Friend,
and we have for the measurement basis of super-Wigner $W$ the following (relevant)  basis states of the joint system $s\otimes F$:
    \begin{eqnarray} \label{F2-final}
    \ket{j^{(2)}; W_1}_{sF} = \sum_{k=1,2} \braket{k^{(3)}}{j^{(2)}}\ket{k^{(3)}}\ket{k^{(3)}}_{F}
    \end{eqnarray}
    Both in (\ref{PSI}) and (\ref{PSI2}) we  have the same unitary matrix $\braket{j^{(2)}}{l^{(1)}}$ which relates  initial basis $n=1$ with the final measurement basis $n=2$. Thus the probability amplitudes are identical.}

    The consequence of this is that the counterfactuality of outcomes of virtual-Friends, defined by what would have been factually observed had this been an actual full measurement, forces the probabilities of the counterfactual outcomes to follow exactly the quantum predictions for concurrent factual measurements. E.g. $P(r,s, t|n, n',n'')_{W_1,F_2,F_3}$ is identical with $P(r,s,t|n, n',n'')_{W_1,W_2,W_3} $, where $r,s,t=\pm1$ are the outcomes, $n,n',n''=1,2,3$ are measurement settings at stations $m=1,2,3$, respectively, and $Z_1,Z_2,Z_3$, where $Z_m=W_m$ or $F_m$, denote the agent performing the measurement or pre-measurement at station $m$.
    
    Thus,   counterfactual outcomes by virtual Friends, if they are in  a configuration of settings which allows perfect GHZ correlations, must also reveal such a correlation.
    
    In the case of the GHZ state (\ref{GHZ}) and local measurements in bases:
    \begin{equation}
         \ket{\pm1, \phi_m}=\frac{1}{\sqrt{2}}\left(\ket{0^{(1)}}_{s_m}\pm e^{i\phi_m}\ket{1^{(1)}}_{s_m}\right),
    \end{equation}
   for observables $\Hat{O}_m(\phi_m)=\sum_{r=\pm1}r\ket{r, \phi_m}_{s_ms_m}\bra{r, \phi_m} $ the correlation function
    $E(\phi_1,\phi_2,\phi_3)$ reads: $\langle\otimes_{m=1}^3 \Hat{O}_m(\phi_m)\rangle _{GHZ}=\cos{(\sum_{m}\phi_m)}$. 
    Phase setting $\phi_m=0$ gives basis $\ket{l^{(2)}}_{s_m}$, and $\phi_m=\pi/2$ gives basis $\ket{l^{(3)}}_{s_m}$.
    For specific local phases we have GHZ perfect correlations, e.g. 
    \begin{eqnarray}\label{GHZ-corr}
     &E(0,\pi/2,\pi/2)=E(\pi/2,\pi/2,0)&\nonumber \\&=E(\pi/2,0,\pi/2)=-E(0,0,0)=-1,&
    \end{eqnarray}
    
     Recall that super-Wigners measure in a basis which gives the same probability amplitudes as basis $\phi_m=0$, while their virtual-Friends make earlier pre-measurements in basis defined by $\phi_m=\pi/2$. Let us denote the actual results by super-Wigners in a given run of the GHZ experiment by $w_m$, and counterfactual outcomes by virtual Friends  by $f_m$. From (\ref{GHZ-corr}) we have 
    $
       w_1w_2w_3=1, 
    $
   but the counterfactual outcomes of virtual Friends  must also satisfy 
   $
       w_1f_2f_3=
       f_1w_2f_3= 
       f_1f_2w_3=-1. 
   $
   Since  $w_m$ have values $\pm1$, the four above relations lead to
    $(f_{1}f_{2}f_{3})^2=-1$, 
    that is the product of the hypothetical counterfactual pre-measurement `results' is... {\it imaginary}, while it must be $\pm1$. 
    The counterfactual outcomes of virtual-Friends form  an internally inconsistent notion, they do not exist even in theory. $\square$ 
    
    The above reasoning differs from the one of GHZ. We have one factual full measurement situation, and three consistency statements on counterfactual outcomes of virtual-Friends pertaining to the same factual situation, elements of which are also the pre-measurement devices of  virtual Friends. 
    We do not need assumptions of free choice of measurement settings by the agents, as all settings are fixed. Neither we need here the Bell-like locality assumption, based on Einstein's causality. The measurement stations do not have to be spatially separated. 
    All that we assumed is that factual outcomes and counterfactual outcomes have to satisfy mutual relations, which are consistent with quantum mechanical predictions. 
    
    The GHZ-like reasoning by Brukner \cite{BRUKNER}, as well as his Bell-like argumentation,  was to show that there are no ``observer independent facts". If used to invalidate the existence of counterfactual outcomes by virtual Friends, cannot do the job, as it also invokes assumptions of freedom of choice of measurement settings by each of the super-Wigners, and Bell-like locality. Thus the contradictions  by Brukner  refute only the conjunction of these two assumptions and existence of counterfactual outcomes of pre-measurements by virtual Friends. They do not invalidate the existence of the counterfactual values by themselves. Mere acceptance of non-locality leaves the  counterfactual outcomes  a viable notion. 
    
    \marcinN{Recently a new argument against \textit{persistent reality of Wigner's Friend perception} (existence of virtual Friend's outcome of pre-measurement) was presented \cite{BRUKNERNEW}.
    The authors show a divergence between two scenarios: the one with pre-measurement done by (virtual) Friend {(\it essentially, our step two) } and the second with measurement record obtained by Friend followed by an application of the state-update rule specified by the conditioning on the Friend's outcome {(effectively our step one, as the quantum theory of measurement is consistent with state update upon measurement, when both are  applied to the whole statistical  ensemble of equivalently prepared systems described by the initial quantum state)}. We show here in {\it step three} that  counterfactual reality of virtual Friend's `outcomes' is an inherently self-contradictory notion, which {\it does not have even a  transitory reality}.}

\marcin{In Wigner's Friend-type experiments one is left with the following two possibilities of interpreting the action done by the Friend: either it is an absolutely irreversible quantum measurement or it is a reversible pre-measurement to which one cannot ascribe any notion of outcome in logically consistent way. Our findings nullify a family of apparent paradoxes presented in recent years, especially the ones in \cite{FR-RN} and \cite{BRUKNER, EXP1, EXP2}}. 
Quantum mechanics consistently describes its use, provided we follow the  dictum by  Bohr, \cite{BOHR}, quoted in our  Abstract\footnote{we have found this very relevant quotation  in \cite{BRUKNER}.}.
Our no-go theorem for existence of outcomes of pre-measurements is strong argument for objectivity of the decoherence process, as a `measurement', which can be undone by some super-Wigner, cannot have outcomes. 
{This calls for at least a re-consideration of the approach to Wigner's Friend  following \textit{Relational Quantum Mechanics} \cite{ROVELLI}, in which the Authors claim that measurement outcomes are objective only relative to a given observer}.,

\marcinN{Moreover our result invalidates all the attempts to solve the apparent contradictions presented in \cite{FR-RN} and \cite{BRUKNER} by redefining the probability assignment rules for sequential measurements involving pre-measurement outcomes \cite{BAUMANN18, BAUMANN20}. In brief words: one cannot assign any notion of probability to a non-existing 'outcome'.} The dictum by  Peres: \textit{unperformed experiments have no results} \cite{PERES} is now extended to pre-measurements: they also have no results, even if they are performed. {\color{black} One may  argue that the reasoning  in {\it step one}  is a tautological implication, however this emphasizes a solid consistency of the quantum mechanical description of  measurement process, which takes into account all physically significant circumstances, and therefore leads to no paradoxes, although it might be counter-intuitive.}

\marcinA{The implications  might be broader. One can see them as revealing an internal inconsistency of the modified concept of the (virtual) Friend (initiated in \cite{DEUTSCH}), regardless of her nature (whether it is a conscious agent, a detector perhaps coupled with a memory). 
Awareness, or records, of measurement outcomes may require Friends to  be complicated enough to be beyond the Heisenberg cut (see our discussion of this concept in Appendix \ref{ApB}), in an effectively classical realm \cite{ZUREK3} resulting from decoherent uncontrollable dynamics of their zillions of degrees of freedom, and perhaps an influence of external environments. All this is 
in line with the quantum theory of measurement, and thus in the description of Friends decoherence is a necessity. Therefore, experiments with virtual Friends replaced by a single degree of freedom, like \cite{EXP1,EXP2}, just cannot emulate them, and  are not tests of such concepts.  }

{\it Acknowledgments.}---The work is part of the ICTQT IRAP (MAB) project of FNP, co-financed by structural funds of EU. MZ dedicates this work to the memory of prof. dr Fritz Haake.

\appendix

\section*{Appendix}
The first section of Appendix is entirely devoted to our direct analysis of the Frauchiger-Renner gedankenexperiment with the use of the quantum measurement theory. It can be read independently of the main text. This is why there is some overlap with the main text.
The second section briefly discusses the Heisenberg Cut, as the purported subjectivity of the notion, is often used against the quantum measurement theory. 
We show that the Cut, despite being movable, is well defined operationally in any experiment.
Section three shows how decoherence at the level of Friends affects the gedankenexperiment of Brukner.
Section four gives basic ideas about quantum-marker quantum eraser experiments. \marcin{The last section provides a brief description of an experimental arrangement in which a Friend performs a pre-measurement in a locally chosen basis, whereas Wigner decoheres the system without knowing the outcome unless he opens the Friend's lab to see the setting of a pre-measurement.}

\section{Analysis of Frauchiger-Renner gedankenexperiment }
\label{ApA}
\subsection{Simple description of  underlying process: no Wigners, no Friends }
\label{Apa1}
We shall show the gist of the process used by Frauchiger and Renner by mapping it into an interferometric process.
The notation which we shall use will differ from the one in \cite{FR-RN}, as we aim at its simplicity. 

Imagine two entangled qubits, $g$ and $s$. The letters are to remind us about the quantum random number $g$enerator and the $s$ystem used in \cite{FR-RN}.
Assume that they are in the following initial state

\begin{equation} \label{1}
\ket{init}_{gs}=\sqrt{\frac{1}{3}}\ket{H}_g\ket{-1}_s+\sqrt{\frac{2}{3}}\ket{T}_g\ket{+}_s,
\end{equation}
where $\ket{H}_g$ and $\ket{T}_g$,  and states $\ket{\pm1}_s$  form orthonormal bases, and $\ket{\pm}_s=\sqrt{\frac{1}{2}}(\ket{-1}_s\pm\ket{+1}_s)$.
The state can be put in an alternative way as

\begin{eqnarray} \label{2}
\ket{init}_{gs}=\sqrt{\frac{1}{3}}\big[(\ket{H}_g+\ket{T}_g )\ket{-1}_s+\ket{T}_g\ket{+1}_s\big] \nonumber\\
\end{eqnarray}

A glance at the formulas \eqref{1} and \eqref{2} shows us that:
\begin{itemize}
\item 
Situation 1: A measurement of $g$ in a basis consisting of $\ket{\pm}_g=\sqrt{\frac{1}{2}}(\ket{H}_g\pm\ket{T}_g )$, and a measurement of $s$ in basis $|\pm\rangle_s$, according to (\ref{2}), gives "$-$" for $g$ and "$-$" for $s$ with a non-zero probability.
\item
Situation 2: A fully complementary measurement in basis $\{\ket{H}_g, \ket{T}_g\} $,
 and a measurement of $s$ in basis $|\pm\rangle_s$, according to (\ref{1}), cannot lead to $T$ for $g$ and "$-$" for $s$.
 That is, the joint probability of $T$ for $g$, and "$-$" for $s$ is zero.
\end{itemize}

Only the first itemized situation, as we shall see further on, allows``$\bar{w}=\bar{ok}$ and $w=ok$"  `halting condition'  of \cite{FR-RN} (i.e. results "$-$" for both $g$ and $s$), which is the one on which the authors of \cite{FR-RN}  base their paradox.
In the second, complementary, situation $T$ and $w=ok$ is impossible.

\subsection{Reworded original description involving pre-measurements by Friends}
\label{ApA2}
Let us now introduce agents in the form of (super) Wigners $W_g$ and $W_s$, and their respective (virtual) Friends (see footnote \footnote{Super Wigners because they have powers not envisaged by E.P. Wigner \cite{WIGNER}, and virtual Friends, because they would preform only pre-measurements, and not measurements as it was assumed by E.P. Wigner}),
 $F_g$ and $F_s$, as well as their devices $D_g$ and $D_s$.  All this is with the obvious relation to the two laboratories of the Friends: $L_g$, which supposedly generates inside a random output of a quantum coin $H=heads$ and $T=tails$, and $L_s$, the one which operates on the system $s$ emitted by $L_g$ in correlation with $H$ or $T$ (pre-measurement) `result'.  

The initial state of the `coin' of the generator is $\sqrt{\frac{1}{3}}\ket{H}_g+\sqrt{\frac{2}{3}}\ket{T}_g$. The internal dynamics of $L_g$ leads to the following entangled state:
\begin{equation} \label{3}
\ket{init}_{L_gs}=\sqrt{\frac{1}{3}}\ket{H}_{L_g}\ket{-1}_s+\sqrt{\frac{2}{3}}\ket{T}_{L_g}\ket{+}_s, 
\end{equation}
where in our notation $\ket{X}_{L_g}=\ket{X}_{g} \ket{X}_{D_g}\ket{X}_{F_g}$, and $X=H,T$. The kets related to the subsystems $g$, $D_g$ and $F_g$ are these which are correlated with each other in the pre-measurement process, i.e. in fact we have:
\begin{eqnarray}
\ket{init}_{L_gs}=\sqrt{\frac{1}{3}}
\ket{H}_{g} \ket{H}_{D_g}\ket{H}_{F_g}
\ket{-1}_s \nonumber\\
+\sqrt{\frac{2}{3}}
\ket{T}_{g} \ket{T}_{D_g}\ket{T}_{F_g}
\ket{+}_s.
\end{eqnarray}

The system $s$ is sent to lab $L_s$. As the result of a \textit{pre}-measurement by $F_s$,  with respect to the basis $\ket{\pm}_s$, one gets the following overall state:
\begin{equation}\label{5}
\ket{init}_{L_gL_s}=\sqrt{\frac{1}{3}}\ket{H}_{L_g}\ket{-1}_{L_s}+\sqrt{\frac{2}{3}}\ket{T}_{L_g}\ket{+}_{L_s}, 
\end{equation}
where $\ket{+}_{L_s}=\sqrt{\frac{1}{2}}(\ket{-1}_{L_s}+\ket{+1}_{L_s})$.
This state  is isomorphic with (\ref{1}).
Equivalently:
\begin{equation}
\ket{init}_{L_gL_s}=\sqrt{\frac{1}{3}}\big[(\ket{H}_{L_g}+\ket{T}_{L_g} )\ket{-1}_{L_s}+\ket{T}_{L_g}\ket{+1}_{L_s}\big],
\end{equation}
where in our notation $\ket{Y}_{L_s}=\ket{Y}_{s} \ket{Y}_{D_s}\ket{Y}_{F_s}$, with $Y=\pm1$.
Please note that it is isomorphic with (\ref{2}). 

Now enter the (super) Wigners, who make full proper measurements on the respective labs. 
$W_g$  measures in a basis containing $\ket{\pm}_{L_g}=\sqrt{\frac{1}{2}}(\ket{H}_{L_g}\pm\ket{T}_{L_g})$,
whereas $W_s$ in a basis containing $\ket{\pm}_{L_s}$.  Whenever $W_g$ obtains $\ket{-}_{L_g}$ there is probability $1/2$ that $W_s$ obtains $\ket{-}_{L_s}$. 
This is the $ok$ and $\bar{ok}$ situation of \cite{FR-RN}.

Despite the fact that within $L_g$ only a (reversible) pre-measurement takes place, it is assumed in \cite{FR-RN} that somehow $F_g$ may think that a real measurement took place with a definite result $H$ or $T$ (this is shown further on here to be an internally  inconsistent idea). When one considers $T$ as a possible state of mind of $F_g$, then to $F_g$ `ok' by $W_s$ is (seems?) impossible (see the previous section, situation 2,  and recall the isomorphism). This is because in such a situation system $s$ is sent to $L_s$ in state $\ket{+}_s$ which after the pre-measurement interaction in $L_s$ leads to $\ket{+}_{L_s}$, while `ok' of $W_s$ is associated with  $\ket{-}_{L_s}$. The purported paradox is that, according to the Authors, one can have $T$ for $F_g$ as her `outcome', while both Wigners obtain outcomes "$-$" ($ok$ and $\bar{ok}$).

Note that the initial state (\ref{3}) is not stated openly in \cite{FR-RN} as the crux of the protocol, but only mentioned in passing, during the discussion. When the main protocol is described, it is  said that whenever $F_g$ sees $H$,  system $s$ in state $\ket{-1}_s$ is sent, whereas for $T$ it is sent in $\ket{+}_s$. Still the description using (\ref{3}) is the only one consistent with  the situation resulting with $w=ok$ and $\bar{w}=\bar{ok}$ (which is isomorphic with situation 1). An incoherent  mixture of 
$
\ket{H}_{g} \ket{H}_{D_g}\ket{H}_{F_g}
\ket{-1}_s 
$ and
$
\ket{T}_{g} \ket{T}_{D_g}\ket{T}_{F_g}
\ket{+}_s
$ in the case of $F_g$ knowing that $T$ happened leads to an isomorphism with situation 2, and  one cannot have   $w=ok$ and $\bar{w}=\bar{ok}$.


\subsection{Description involving full proper quantum measurements by Friends}
\label{ApA3}
Here we shall consider the consequences of using the full quantum measurement theory, which involves decoherence and irreversibility of recorded results. Note that this means that we assume that quantum theory has a `universal validity' which encompasses the full measurement process.  We assume here that both Friends and Wigners perform full proper measurements.

Consider the stages of the measurement. Let us take lab $L_g$ which is slightly more important here, but a similar description can be used in the case of the other lab.
Assume the system $g$ is in a superposition $\alpha\ket{H}_g+\beta\ket{T}_g$.  In the pre-measurement stage it gets entangled with the pointer variable of device $D_g$. 
We get as a result 
$
\alpha\ket{H}_g\ket{H}_{D_g}+\beta\ket{T}_g\ket{T}_{D_g}.
$

Now enters the decoherence. The environment might be in a mixed initial state, but for simplicity of presentation we shall write it down as a pure one.
Within the considered scenario the decoherence takes place due to interaction with the environment constrained to  lab $L_g$. The two distinguishable states of the environment related with distinct states of the pointer-device $X=H, T$
will be denoted as $\ket{X}_{E_g}$.
Thus if one additionally takes into account the emitted system $s$, the state of lab $L_g$ plus $s$ is, after the full measurement process, given by:
\begin{eqnarray} 
\alpha\ket{H}_g\ket{H}_{D_g}\ket{H}_{F_g}\ket{H}_{E_g}\ket{-1}\nonumber \\
+\beta\ket{T}_g\ket{T}_{D_g}\ket{T}_{F_g}\ket{T}_{E_g}\ket{+1}.
\end{eqnarray}

The trouble is that $W_g$ upon his measurement, which perhaps could involve the imaginary conscious Friend $F_g$, cannot undo the irreversible interaction with the environment (and thus cannot undo the records). His power is limited to measurements of $g$, maybe even $g\otimes D_g$, and if he is cruel and super-sophisticated to $g\otimes D_g\otimes F_g$. But in all such cases because of the entanglement with environment $E_g$, effectively, he deals with a reduced density matrix of the subsystems, which is a classically correlated  state (for $\alpha\beta\neq 0$): 
$\ket{H}_g\ket{H}_{D_g}\ket{H}_{F_g}$, with probability $|\alpha|^2,$ and $\ket{T}_g\ket{T}_{D_g}\ket{T}_{F_g}$, with  probability $|\beta|^2$.

Thus a consistent use of the full measurement theory nullifies the paradox.  $F_g$ records either $H$ or $T$. If it is $T$, there is an isomorphism with situation 2, and there is no way to formulate the paradox of \cite{FR-RN}.

 However, all that  leaves us with the task of analyzing the situation of the previous subsection in which Friends perform only {\it pre-measurements}.
We shall show that in such a case the Friends cannot know `outcomes'. This also nullifies the paradox at a deeper level. Therefore we come back to the dictum by  Peres: \textit{unperformed experiments have no results} \cite{PERES}.  One should here add: pre-measurements are not performed measurements as they can be undone.

\section{Operationally defined Heisenberg Cut}
\label{ApB}
Super-Wigners cannot exist without virtual-Friends. The division between them is related to the Heisenberg Cut. The Cut is movable, as with technological progress Super-Wigners may be able to control more and more of degrees of freedom associated with ever more complex Friends. The cut is an objective division between the controllable and uncontrollable part of the full measurement process involving $F_g$ and $W_g$. Still it is defined by the technology used, it is dependent on `all relevant features of the  experimental arrangement'.

Let us attempt to give an operational method to define such a Cut in the experiment considered here. Obviously sealing in a laboratory of our friend (say, Harald) and expecting him to get entangled in  a pre-measurement superposition like  $\ket{\pm}_{L_g}=\frac{1}{\sqrt{2}}\big(\ket{H}_g\ket{H}_{D_g}\ket{H}_{F_g} \pm\ket{T}_g\ket{T}_{D_g}\ket{T}_{F_g}\big)$, seems out of question.
This is because states $\ket{Y}_{F_g}$ are in fact states of very many, say $10^{40}$ or something, degrees of freedom.
Thus we can start our experiments with a qubit Friend, $F_1$, of states spanned by $\ket{X}_{F_1}$. Control and observation by Wigner of superpositions $\frac{1}{\sqrt{2}}\big(\ket{H}_g\ket{H}_{D_g}\ket{H}_{F_1} + e^{i\phi}\ket{T}_g\ket{T}_{D_g}\ket{T}_{F_1}\big)$ is very easy, as it is enough to impose a phase shift on system $g$. Here we assume that $D_g$, the pointer variable, is also a qubit. However Winger may try a more complicated system, with an artificial Friend composed of $m$ qubits. Thus we replace $\ket{Y}_{F_1}$ by $\ket{Y}_{\vec{F}}=\otimes_{n=1}^m\ket{Y_n}_{F_n}$, and require that ${}_{\vec{F}}\bra{T}\ket{H}_{\vec{F}}=0$. To get interference it is again enough to phase shift the qubit $g$  to get 
$\frac{1}{\sqrt{2}}\big(\ket{H}_g\ket{H}_{D_g}\ket{H}_{\vec{F}} + e^{i\phi}\ket{T}_g\ket{T}_{D_g}\ket{T}_{\vec{F}}\big)$  and then for example to unitarily transform $\ket{T}_g\ket{T}_{D_g}\ket{T}_{\vec{F}}$ into $\ket{T}_g\ket{H}_{D_g}\ket{H}_{\vec{F}}$, to get $\frac{1}{\sqrt{2}}\big(\ket{H}_g + e^{i\phi}\ket{T}_g\big)\ket{H}_{D_g} \ket{H}_{\vec{F}}$. Finally Wigner makes his own measurement of $g$ in basis $\frac{1}{\sqrt{2}}(|H\rangle_g\pm|T\rangle_g)$. This will lead to a perfect interference. Our experimenter may be ambitious, and would aim at the biggest possible $m$. At a certain $m$ his ability to transform $\ket{T}_{\vec{F}}$ into $\ket{H}_{\vec{F}}$, i.e to control the microscopic state of Friend will be lost. Possible reasons may be the following: too high $m$ to handle, or Friend's interaction with the environment (uncontrollable by definition). Note that the environment can be thought of as a part of the Friend, hence this case is reduced to a question of a too high $m$ for a quantum  control of the Friend's degrees of freedom. The threshold $m_{th}$ signifies the Heisenberg cut. Above it, the state of the system $g$, pointer, and Friend is a probabilistic mixture of 
$\ket{H}_g\ket{H}_{D_g}\ket{H}_{\vec{F}}$ and  $\ket{T}_g\ket{T}_{D_g}\ket{T}_{\vec{F}}$. This signifies that Friend effectively makes a full measurement (an effective decoherence steps in), and a classical description of the Friend is permitted \cite{BOHM}.

To put it short, if Wigner is able to obtain a $\phi$-dependent  interference in an experiment on a triple: system, plus pointer, plus Friend in state $\ket{\phi}_{L_g}=\frac{1}{\sqrt{2}}\big(\ket{H}_g\ket{H}_{D_g}\ket{H}_{F_g} +e^{i\phi}\ket{T}_g\ket{T}_{D_g}\ket{T}_{F_g}\big)$ by making measurements in the basis $\ket{\pm}_{L_g}=\frac{1}{\sqrt{2}}\big(\ket{H}_g\ket{H}_{D_g}\ket{H}_{F_g} \pm\ket{T}_g\ket{T}_{D_g}\ket{T}_{F_g}\big)$, then his Friend performs only a pre-measurement. System, plus pointer plus Friend can and should be described using quantum mechanics. If such interference is impossible, quantum mechanical description is still possible, however it is redundant. We have passed the Cut.

\marcin{One often encounters an argument that the objectivity of the Heisenberg Cut holds only \textit{For All Practical Purposes - FAPP}, and therefore is not fundamentally justified.
However, as pointed out in \cite{STREATER}, if we accept this viewpoint, we have the same problem in classical physics in the theory of phase transitions. as formally phase transitions occur in the thermodynamic limit.  Formally the quantum measurement is absolutely irreversible only in the limit of infinite size of the device (or its environment). In our opinion both of these processes are fundamentally objective, however their justification needs an idealization in the mathematical description. Therefore the problem is not with the physical processes, but rather with the understanding that the formalism itself is only a tool to describe them, and  some idealizations within the formalism are unavoidable.}

\section{Gedankenexperiment of Brukner with decoherence of pre-measurements by Friends}
\label{ApC}

The initial state is a maximally entangled state, say singlet,  of two qubits.
 After a measurement process  by Friends the statistical ensemble of the experiment is described by a  mixture of product states which are possible results of the joint measurement by Friends in a factorizable basis, which obviously is a separable state, say $\varrho_{sep}$. Thus if Friends indeed measure, not just pre-measure, one cannot violate any Bell-like inequality, which involves as the second stage also measurements by Wigners, because in all formulas one can replace the singlet by $\varrho_{sep}$.
 
 Decoherence process is objective, and it is necessary to turn initial system-pointer correlation into measurement results, which are facts. Quantum mechanics does not predict which fact is bound to happen in a given run. Like any probabilistic theory it  precisely determines probability distribution of facts for the whole statistical ensemble of equivalent preparations and equivalent observations.

\section{Quantum marker - quantum eraser experiments}
\label{ApD}

A beautiful example of  an experiment, which undoes a pre-measurement,  can be found in \cite{KWIAT1, KWIAT2, KWIAT3}.
Here we shall present just the basic theory of such experiments, which is not a direct description of the experiments in \cite{KWIAT1, KWIAT2, KWIAT3}.

Imagine a Mach Zehnder interferometer with the front beam-splitter replaced by a polarizing beamsplitter which splits polarizations $h$, horizontal and $v$, vertical.  A photon of polarization  $|d\rangle = \frac{1}{\sqrt{2}}(|h\rangle+|v\rangle)$  enters the beamsplitter, and leaves it in a state
 $ \frac{1}{\sqrt{2}}(|h\rangle|1\rangle+|v\rangle|2\rangle)$, where $\ket{j}$, with $j=1,2$, denotes the photon being in the upper, 1, or lower, 2,  arm of the interferometer. Let us place a phase shifter in the upper arm. After that the state becomes $ \frac{1}{\sqrt{2}}(e^{i\phi}|h\rangle|1\rangle+|v\rangle|2\rangle)$.
Let the exit beamsplitter be, as it is usually the case, a symmetric polarization neutral one. Thus it transforms $\ket{1}$ into $ \frac{1}{\sqrt{2}}(|1\rangle+i|2\rangle)$
and $\ket{2}$ into $ \frac{1}{\sqrt{2}}(|2\rangle+i|1\rangle)$. The final state is thus: 
\begin{equation}
 \frac{1}{{2}}\left(e^{i\phi}|h\rangle|1\rangle+ie^{i\phi}|h\rangle|2\rangle +|v\rangle|2\rangle+i|v\rangle|1\rangle\right). \end{equation}
We have a superposition of four orthogonal states. The probability of finding the photon in exit beam $1 $ is $1/2$. We have no interference (no dependence on $\phi$). This is because we do not have the required indistinguishability of paths `taken' by the photon within the interferometer. The polarizations mark the paths, or if you like (pre) measure them (well, it is the polarization beamsplitter that does this).  However this can be undone, by placing in front of the detector a polarization filter letting through only polarization $\ket{d}=\frac{1}{\sqrt{2}}(\ket{h}+\ket{v})$. 
Then the amplitude for the photon to reach  the detector in say beam 1 would be  $ (\frac{1}{\sqrt{2}})^3(e^{i\phi} + i)$, which gives an interference of visibility  equal to 1. The paths are indistinguishable beyond the polarization filter. Obviously if we place some photon-non-destroying detection devices in the internal paths of the interferometer recovering of the interference would be impossible.

\section{A trivial operational example of situation in which ``Friend" performs a pre-measurement, informs Wigner that it has been done, but does not reveal the outcome. The decoherence is done by Wigner}
\label{ApE}

\marek{As we have shown that `results' of pre-measurement do not exist, the contraption that we shall present will not have the feature that Friend "sees a definite outcome". However it is a feasible realization of Deutsch's idea in all other respects, and can be used in demonstration of the kind presented in \cite{EXP1} and \cite{EXP2} in which we have {\it bona fide} Friends.

Via an optical fiber, or whatever, we feed single photons into Friend's sealed lab. She is an automaton which controls a polarization rotator and a device (set of wave plates) which changes phases between two orthogonal linear polarizations.   This is in front of one of the entry ports of a polarization beamsplitter distinguishing say $h$ and $v$ polarizations. 

Thus, the lab is equipped with a universal polarization beamsplitter. It directs two orthogonal polarizations of light into two different exit beams. Namely, an operational method of obtaining such a beamsplitter is to use a standard $h$ and $v$ polarization separating beamsplitter, and to place in front of its (used) input port a suitable polarization rotator, and an additional tunable/replacable wave-plate which changes the relative phase of $h$ and $v$ polarizations. Such a device is capable to perform any U(2) transformation on polarization states. If one sets transformation $U^{\dagger}(\alpha, \phi)$   to be such that  it transforms  $\ket{h} $ into $\cos{\alpha}\ket{h}+ e^{i\phi}\sin{\alpha}\ket{v}$, and $\ket{v} $ into $-e^{i\phi}\sin{\alpha}\ket{h}+\cos{\alpha}\ket{v}$, the full device is such that photons of elliptic polarization {$U(\alpha, \phi)\ket{h}$} exit by output $h$ (with polarization $h$) and those of polarization 
 $U(\alpha, \phi)\ket{v}$ via exit $v$ (with polarization $v$). 

Additionally let us assume that both exit beams of the universal polarizer are at an angle with the original beam. Thus when there is no polarizer the photons travel along the original path (this can be done in a more sophisticated way than by removing the polarizer - by putting in front of it a Mach-Zehnder interferometer, and allowing in it two internal phase differences, one that directs all light to an exit port which leads to the polarizer and another one which directs the light to the other exit port, which does not lead to the polarizer).

Let us now (partially) seal the lab, in such a way that the local  Wigner
has no knowledge which two elliptic polarizations are split by the beamsplitter (simply, he does not know the local $U(\alpha, \phi)$). That is we put the automaton-Friend and her universal polarization beamsplitter into a black box of sorts. Still we would allow her lab (box) to have three exit optical fibers, two with their inputs behind the beamsplitter, and one for the "no-measurement" channel. If Wigner registers a photon in the last channel, then this signals that no pre-measurement was done. Registration in the other two channels signals that a pre-measurement was done within the lab (for simplicity we assume a perfect detection efficiency).
No matter what is the device  setting by Friend (the choice of $U(\alpha, \phi)$), when she pre-measures, the final result of this is 
a detection of a photon  by Wigner in one of the "measurement" exits (their polarization, or presence  in this or other exit, does not reveal what was the setting by Friend). 
Until Wigner unseals the box-lab to look inside at the setting of the device,  he does not know which pre-measurement was done. Still, as it is him who uses macroscopic detectors which send the record to, say a computer, it is him who introduces the decoherence. 

Of course, this contraption can be expanded in such a way that the detection stations have some mechanism to "store" the photons in order to delay the moment of detection, e.g. in a form of an excitation of an atom in a trap, or some other means which are currently becoming more and more feasible. In such a case Wigner can learn the Friend's setting, before the final decoherence in his detection system.}


\begin{thebibliography}{2}%
\makeatletter
\providecommand \@ifxundefined [1]{%
 \@ifx{#1\undefined}
}%
\providecommand \@ifnum [1]{%
 \ifnum #1\expandafter \@firstoftwo
 \else \expandafter \@secondoftwo
 \fi
}%
\providecommand \@ifx [1]{%
 \ifx #1\expandafter \@firstoftwo
 \else \expandafter \@secondoftwo
 \fi
}%
\providecommand \natexlab [1]{#1}%
\providecommand \enquote  [1]{``#1''}%
\providecommand \bibnamefont  [1]{#1}%
\providecommand \bibfnamefont [1]{#1}%
\providecommand \citenamefont [1]{#1}%
\providecommand \href@noop [0]{\@secondoftwo}%
\providecommand \href [0]{\begingroup \@sanitize@url \@href}%
\providecommand \@href[1]{\@@startlink{#1}\@@href}%
\providecommand \@@href[1]{\endgroup#1\@@endlink}%
\providecommand \@sanitize@url [0]{\catcode `\\12\catcode `\$12\catcode
  `\&12\catcode `\#12\catcode `\^12\catcode `\_12\catcode `\%12\relax}%
\providecommand \@@startlink[1]{}%
\providecommand \@@endlink[0]{}%
\providecommand \url  [0]{\begingroup\@sanitize@url \@url }%
\providecommand \@url [1]{\endgroup\@href {#1}{\urlprefix }}%
\providecommand \urlprefix  [0]{URL }%
\providecommand \Eprint [0]{\href }%
\providecommand \doibase [0]{http://dx.doi.org/}%
\providecommand \selectlanguage [0]{\@gobble}%
\providecommand \bibinfo  [0]{\@secondoftwo}%
\providecommand \bibfield  [0]{\@secondoftwo}%
\providecommand \translation [1]{[#1]}%
\providecommand \BibitemOpen [0]{}%
\providecommand \bibitemStop [0]{}%
\providecommand \bibitemNoStop [0]{.\EOS\space}%
\providecommand \EOS [0]{\spacefactor3000\relax}%
\providecommand \BibitemShut  [1]{\csname bibitem#1\endcsname}%
\let\auto@bib@innerbib\@empty
\bibitem [{Note1()}]{Note1}%
  \BibitemOpen
  \bibinfo {note} {We have found this very relevant quotation in \cite
  {BRUKNER}.}\BibitemShut {Stop}%
\bibitem [{Note2()}]{Note2}%
  \BibitemOpen
  \bibinfo {note} {Super Wigners because they have powers not envisaged by E.P.
  Wigner \cite {WIGNER}, and virtual Friends, because they would preform only
  pre-measurements, and not measurements as it was assumed by E.P.
  Wigner}\BibitemShut {NoStop}%
\end{thebibliography}%


\begin{thebibliography}{99}

\bibitem{FR-RN}
 D. Frauchiger and R. Renner, Nat. Comm. \textbf{9}, 3711 (2018).
 
 \bibitem{GHZ}
 D. M. Greenberger, M. A. Horne, and A. Zeilinger in  {\it Bell’s theorem, Quantum Theory, and Conceptions of the Universe}, M. Kafatos (Ed.), (Kluwer, Dordrecht, 1989), p. 69-72.
 
 
\bibitem{BRUKNER}
C. Brukner, Entropy \textbf{20}, 350 (2018)

 \bibitem{EXP1}
M. Proietti, A. Pickston, F. Graffitti, P. Barrow, D. Kundys, C. Branciard, M. Ringbauer, and A. Fedrizzi,
 Science Advances \textbf{20} (5), 9, eaaw9832 (2019).

\bibitem{EXP2}
K.-W. Bong, A. Utreras-Alarc\'on, F. Ghafari, Y.-Ch. Liang, N. Tischler, E. G. Cavalcanti, G. J. Pryde and H. M. Wiseman,  Proc. SPIE 11200, AOS Australian Conference on Optical Fibre Technology (ACOFT) and Australian Conference on Optics, Lasers, and Spectroscopy (ACOLS) 2019, 112001C (2019); Nature Physics \textbf{16}, 1199–1205 (2020).

\bibitem{BOHM} 
D. Bohm, \textit{Quantum Theory}, Prentice-Hall, New York, (1951).

\bibitem{BUSCH} 
P. Busch, P. J. Lahti and P. Mittelstaedt, \textit{The Quantum Theory of Measurement
}, Springer-Verlag, Berlin Heidelberg, (1996).

\bibitem{HAAKE}
F. Haake and M. \.Zukowski, Phys. Rev. A \textbf{47}, 2506 (1993).

\bibitem{ZUREK1}
W. H. Zurek, Phys. Rev D \textbf{26}, 1862 (1982).

\bibitem{ZUREK2}
W. H. Zurek, Rev. Mod. Phys. \textbf{75}, 715 (2003), see also M. Schlosshauer, Rev. Mod. Phys. {\bf 76}, 1267 (2005). 

\bibitem{OLLIVIER}
H. Ollivier, D. Poulin, and W. H. Zurek, Phys. Rev. Lett. \textbf{93}, 220401 (2004).

\bibitem{KORBICZ}
R. Horodecki, J. K. Korbicz, and P. Horodecki, Phys. Rev. A \textbf{91}, 032122 (2015).

\bibitem{FEYNMAN}
R. P. Feynman, R. B. Leighton, and M. L. Sands, \textit{The Feynman Lectures on Physics}, Volume III, Addison-Wesley Pub. Co. (1965), chapter 5.

\bibitem{KWIAT1}
P. G. Kwiat, A. M. Steinberg, and R. Y. Chiao, Phys. Rev. A \textbf{45}, 7729 (1992).

\bibitem{KWIAT2}
T. J. Herzog, P. G. Kwiat, H. Weinfurter, and A. Zeilinger, Phys. Rev. Lett. \textbf{75}, 3034 (1995).

\bibitem{KWIAT3}
P. D. D. Schwindt, P. G. Kwiat, and B.-G. Englert, Phys. Rev. A \textbf{60}, 4285-4290 (1999).

\bibitem{DREZET18}
A. Drezet, arXiv:1810.10917 [quant-ph] (2018).

\bibitem{FORTIN19}
S. Fortin and O. Lombardi, arXiv:1904.07412 [quant-ph] (2019).

\bibitem{LOSADA19}
M. Losada, R. Laura, and O. Lombardi, Phys. Rev. A \textbf{100}, 052114 (2019).

\bibitem{RELANO1}
A. Rela\~{n}o, arXiv:1810.07065 [quant-ph] (2018).

\bibitem{RELANO2}
A. Rela\~{n}o, Phys. Rev. A \textbf{101}, 032107 (2020).

\bibitem{HEALEY18}
R. A. Healey, Found. Phys. \textbf{48}, 1568–1589 (2018).

\bibitem{SUDBERY19}
A. Sudbery, Int. J. Quant. Found. \textbf{5}, 98-109 (2019).

\bibitem{LALOE18}
F. Lalo\"e, 	arXiv:1802.06396 [quant-ph] (2018).

\bibitem{TAUSK}
D. V. Tausk, 	arXiv:1812.11140 [quant-ph] (2018).

\bibitem{SUAREZ}
A. Suarez, arXiv:1906.10524 [quant-ph] (2019).


\bibitem{WAAIJER}
M. Waaijer and J. van Neerven, arXiv:1902.07139 [quant-ph] (2019).

\bibitem{ROVELLI}
A. Di Biagio, C. Rovelli, arXiv:2006.15543 [quant-ph] (2020).

\bibitem{CAVALCANTI20}
E. G. Cavalcanti, 	arXiv:2008.05100 [quant-ph] (2020).

\bibitem{DEBROTA20}
J. B. DeBrota, Ch. A. Fuchs and R. Schack, arXiv:2008.03572 [quant-ph] (2020).

\bibitem{BUB19}
J. Bub, \textit{Two dogmas' redux} in Quantum, Probability, Logic, pp. 199-215, Springer (2020).

\bibitem{FUCHS-PERES}
C. Fuchs and A.Peres, Physics Today \textbf{53} (3), 70 (2000).

\bibitem{WIGNER}
E. P. Wigner, \textit{Remarks on the mind-body question}, in: I. J. Good, \textit{The Scientist Speculates}, London, Heinemann  (1961).

\bibitem{DEUTSCH}
D. Deutsch, Int. J. Th. Phys. \textbf{24}, 1-41 (1985).

\bibitem{BRUKNERNEW}
P. A. Guérin, V. Baumann, F. Del Santo and \v{C}. Brukner, 	arXiv:2009.09499 [quant-ph] (2020).

\bibitem{PERES}
A. Peres, Am. J. Phys. \textbf{46}(7), 745 (1978).

\bibitem{ENT-SWAP}
M. \.Zukowski, A. Zeilinger, M.A. Horne, A.K. Ekert,
Phys. Rev. Lett. \textbf{ 71}, 4287 (1993).

\bibitem{HARALD}
W Rosenfeld, D Burchardt, R Garthoff, K Redeker, N Ortegel, M Rau, H. Weinfurter,
Phys. Rev. Lett., \textbf{ 119}, 010402 (2017).

\bibitem{ZZW}
M. \.Zukowski, A. Zeilinger, H.Weinfurter,
Ann. NY Acad. Sci. \textbf{ 755}, 91-102 (1995)

\bibitem{BOHR} 
N. Bohr, \textit{Quantum Physics and Philosophy: Causality and Complementarity}, in \textit{Philosophy in Mid-Century, A. Survey}, ed. R. Klibansky, (La Nuova Italia Editrice, Florence, 1958).

\bibitem{BAUMANN18}
V. Baumann and S. Wolf, Quantum \textbf{2}, 99 (2018).

\bibitem{BAUMANN20}
V. Baumann, F. Del Santo, A. R. H. Smith, F. Giacomini, E. Castro-Ruiz and C. Brukner, 	arXiv:1911.09696 [quant-ph] (2019).

\bibitem{ZUREK3}
W.H. Zurek, Physics Today {\bf 44}, 36 (1991).

\end{thebibliography}
\end{document}